\documentclass[prb,twocolumn,hidepacs,preprintnumbers,amsmath,amssymb,superscriptaddress]{revtex4}

\usepackage{graphicx}% Include figure files
\usepackage{dcolumn}% Align table columns on decimal point
\usepackage{bm}% bold math

\begin{document}

\title{Observation of chiral quantum-Hall edge states in graphene}

\author{Dong-Keun Ki}
\affiliation{Department of Physics, Pohang University of Science
and Technology, Pohang 790-784, Republic of Korea}
\author{Sanghyun Jo}
\author{Hu-Jong Lee}
\email{hjlee@postech.ac.kr} \affiliation{Department of Physics,
Pohang University of Science and Technology, Pohang 790-784,
Republic of Korea}\affiliation{National Center for Nanomaterials
Technology, Pohang 790-784, Republic of Korea}

\date{\today}

\begin{abstract}
In this study, we determined the chiral direction of the
quantum-Hall (QH) edge states in graphene by adopting simple
two-terminal conductance measurements while grounding different
edge positions of the sample. The edge state with a smaller
filling factor is found to more strongly interact with the
electric contacts. This simple method can be conveniently used to
investigate the chirality of the QH edge state with zero filling
factor in graphene, which is important to understand the symmetry
breaking sequence in high magnetic fields ($\gtrsim$25 T).
\end{abstract}


\maketitle

\begin{figure}[t]
\includegraphics[width=8.5cm]{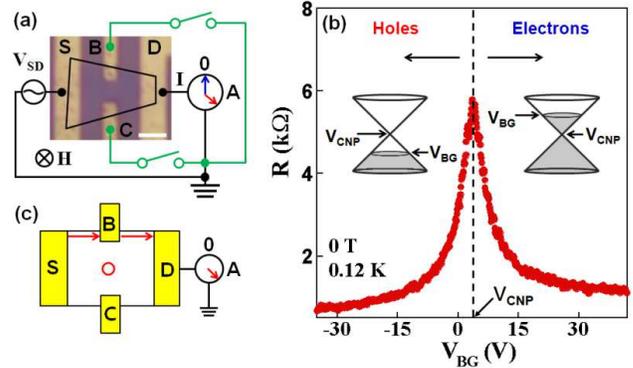}
\caption{\label{fig:sample} (Color online) (a) An optical
microscopic image of the sample and the measurement configuration
(the scale bar represents 2 $\mu$m). The black lines represent the
boundary of the graphene. Switches connected to the EC electrodes
are used to control grounding the electrodes B and C. The current
is measured at the drain D. (b) The BG voltage ($V_{BG}$)
dependence of the resistance in zero magnetic field, which
illustrates a bipolar electric field effect. Inset:
energy-momentum dispersion relations in the hole side;
$V_{BG}$$<$$V_{CNP}$ (left) and in the electron side;
$V_{BG}$$>$$V_{CNP}$ (right). (c) Schematic configuration of the
clockwise QH edge-state circulation in the hole side at positive
$V_{SD}$ with no grounded ECs.}
\end{figure}

\begin{figure}[t]
\includegraphics[width=8.5cm]{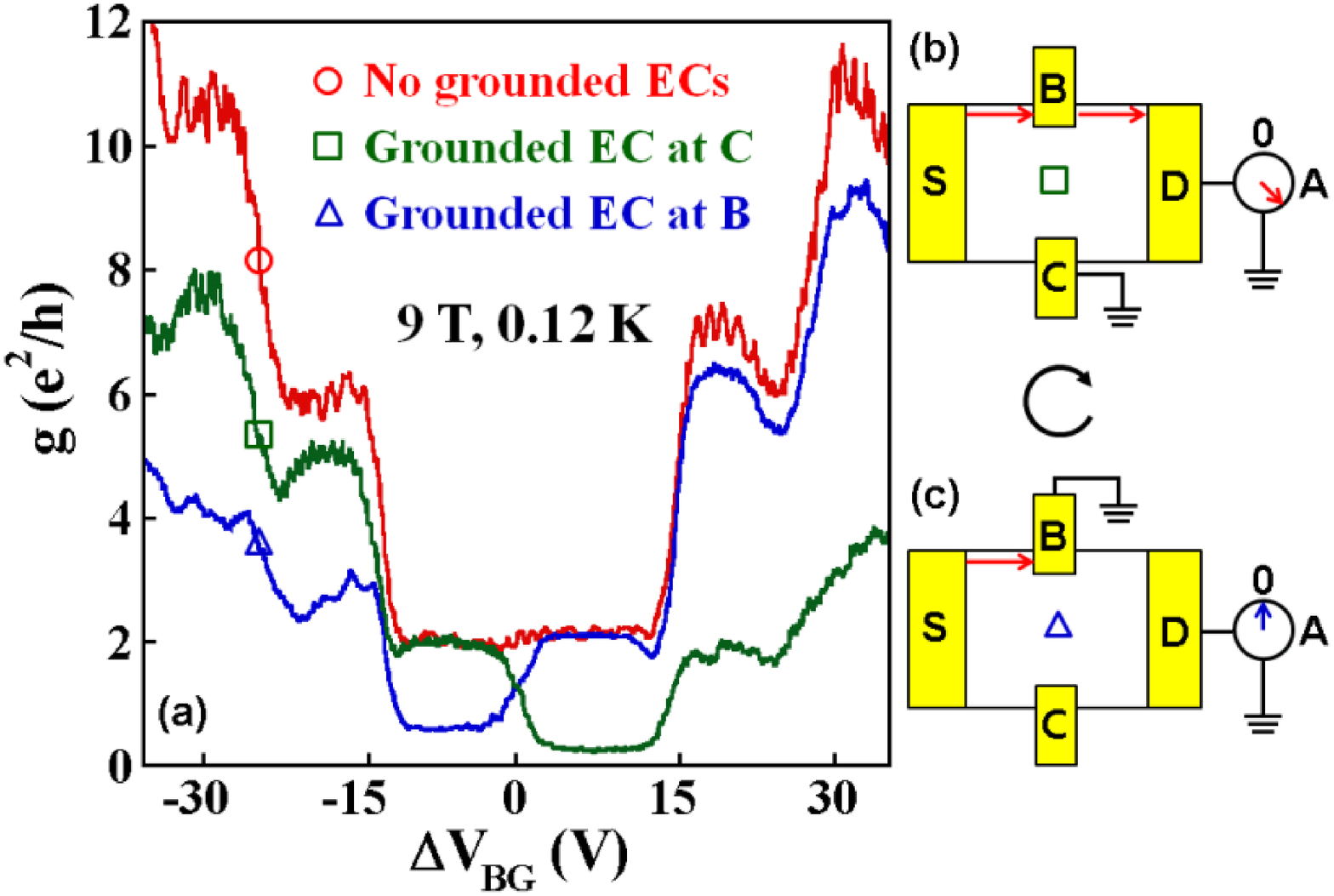}
\caption{\label{fig:datap} (Color online) (a) Two-terminal
conductance ($G$) as a function of a BG voltage difference
($\Delta$$V_{BG}$=$V_{BG}$$-$$V_{CNP}$) in 9 T. Open circle: no
grounded ECs. Open square: the EC grounded at C. Open triangle:
the EC grounded at B. (b,c) Schematic configurations of the
clockwise QH edge state circulation (for hole side and positive
$V_{SD}$) corresponding to the recovery and the reduction of $G$,
respectively, in conductance plateaus.}
\end{figure}

\begin{figure}[t]
\includegraphics[width=8.5cm]{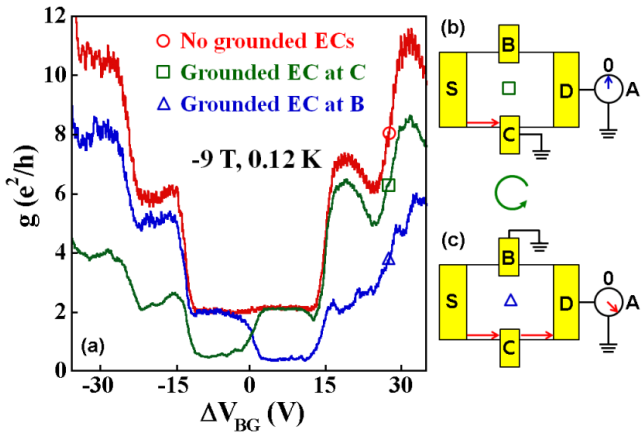}
\caption{\label{fig:datan} (Color online) (a) Two-terminal
conductance ($G$) as a function of a BG voltage difference
($\Delta$$V_{BG}$=$V_{BG}$$-$$V_{CNP}$) in -9 T. Open circle: no
grounded ECs. Open square: the EC grounded at C. Open triangle:
the EC grounded at B. (b,c) Schematic configurations of the
counterclockwise QH edge state circulation (for hole side and
positive $V_{SD}$) corresponding to the reduction and the recovery
of $G$, respectively, in conductance plateaus.}
\end{figure}

At a quantum-Hall (QH) plateau in a two-dimensional (2D) electron
system, conducting electrons in the QH edge states circulate in
one direction (a chiral direction) without any
dissipation.~\cite{Klitzing80,Halperin82,Buttiker88} These QH edge
states can be conveniently used to build quantum devices such as a
solid-state electronic beam splitter.~\cite{Ji03} In graphene, a
honeycomb lattice of $sp^2$-bonded carbon atoms, the relativistic
character of the Dirac-fermionic carriers exhibits the
half-integer QH effect,~\cite{Neto09} which is distinctive from
the integer QH effect in ordinary 2D electronic
systems.~\cite{Klitzing80} It has been generally believed that
carriers in the QH edge states in graphene also circulate in one
direction (one chirality), either clockwise or counterclockwise,
as in a conventional 2D electron gas for a fixed perpendicular
magnetic field.~\cite{Abanin07-2,Williams07,Ozyilmaz07} The chiral
properties of the QH edge states are important, in particular, in
a strong-enough magnetic field as to break either the spin or
pseudospin symmetry, which leads to the QH plateau of zero filling
factor.~\cite{Zhang06,Abanin07,Jiang07,Checkelsky08,Checkelsky09}
Depending on whether the spin or the pseudospin symmetry is broken
first, graphene becomes either ferromagnetic or insulating,
respectively.~\cite{Abanin07-1,Gusynin08} In the QH ferromagnetic
state,~\cite{Zhang06,Abanin07,Jiang07} the QH edge carriers in the
zero plateau are believed to be in a counter-circulating state,
but in the QH insulating state,~\cite{Checkelsky08,Checkelsky09}
the edge states themselves do not exist.

In this study, we examined the chiral direction of the QH edge
states in graphene by using a simple two-terminal transport
measurement scheme while grounding different edge positions of the
sample. Reduction and recovery of the conductance were observed
depending on the position of the grounded edge contacts (ECs), on
the back-gate (BG) voltage ($V_{BG}$), and on the magnetic-field
direction. Our study indicates that the QH edge states in graphene
also possess the chirality,~\cite{Halperin82,Buttiker88,Ki09} and,
moreover, the edge states corresponding to smaller filling factors
are more strongly coupled with the electrical
contacts.~\cite{Krstic08}

Monolayer graphene sheets were mechanically
exfoliated~\cite{Novoselov04} onto a highly doped silicon
substrate covered with a 300-nm-thick oxidation layer, where the
substrate served as the global BG. The device [Fig.
\ref{fig:sample}(a)] was fabricated by the sequential e-beam
lithography and the electrode metallization by
evaporation.~\cite{Ki08} The sample was cooled down to 120 mK in a
dilution fridge (Oxford Instruments, Model Kelvinox), where
electric noises were filtered out by connecting low-pass RC and
$\pi$ filters in series.~\cite{Ki08} At the base temperature, the
two-terminal conductance ($G$=$I$/$V_{SD}$) was measured using the
conventional lock-in technique operated at 13.33 Hz while
monitoring the current with a current preamplifier for the
source-drain bias voltage ($V_{SD}$) of 30 $\mu$V. The measurement
configuration is illustrated in Fig. \ref{fig:sample}(a).

Figure \ref{fig:sample}(b) shows the electric-field control of the
conducting charge carriers,~\cite{Neto09} which reveals a
resistance maximum at $V_{BG}$$\sim$3.8 V. At this charge
neutrality point ($V_{CNP}$), the carriers changes from being
holelike to being electronlike with increasing $V_{BG}$ [see the
inset of Fig. \ref{fig:sample}(b)].~\cite{Neto09} As shown in Fig.
\ref{fig:sample}(b), the resistance asymmetry is present between
the hole and the electron sides, i.e., the mobility ($\mu$) of the
device differs as the carrier type varies. The value of $\mu$ is
$\sim$3600 cm$^2$V$^{-1}$s$^{-1}$ ($\sim$4500
cm$^2$V$^{-1}$s$^{-1}$) for $V_{BG}$=30 V (-20 V). This
electron-hole asymmetry may arise from the unintended chemical
doping of the graphene sheet in the fabrication
process~\cite{Farmer09} or the charge pinning at the interface of
invasive metallic contacts and the graphene sheet.~\cite{Huard08}

Figure \ref{fig:datap}(a) displays the two-terminal conductance
$G$ in units of $G_0$=e$^2$/h as a function of $V_{BG}$ with
respect to $V_{CNP}$ ($\Delta$$V_{BG}$=$V_{BG}$$-$$V_{CNP}$)
measured in $H$=9 T [see Fig. \ref{fig:sample}(a) for the positive
field direction]. The total contact resistance including the
resistance of low-pass RC filters connected in series, $\sim$4.3
k$\Omega$, was obtained independently and subtracted from the
measured two-terminal conductances. With the EC electrodes B and C
being floated, a clear conductance plateau at 2$G_0$ is observed
in the region of $\sim$-15 V$<$$\Delta$$V_{BG}$$<$$\sim$15 V (open
circle). Moreover, the clear plateaus at 6$G_0$ and $\sim$10$G_0$
in the hole side ($\Delta$$V_{BG}$$<$0) indicate the occurrence of
the half-integer QH effect, which confirms the single-layeredness
of our graphene
sheet.~\cite{Williams07,Ozyilmaz07,Abanin08,Williams08} The
distorted $\sim$6$G_0$ and $\sim$10$G_0$ plateaus in the electron
side ($\Delta$$V_{BG}$$>$0) are believed to arise from the reduced
zero-field mobility of our graphene sheet in the electron side
[Fig. \ref{fig:sample}(b)]. Similar distortion of the two-terminal
conductance has been suggested~\cite{Abanin08} and experimentally
confirmed as well.~\cite{Williams08}

As denoted by the open square symbol in Fig. \ref{fig:datap}(a),
the conductance plateau at 2$G_0$, with the EC grounded at the
electrode C, remains unaltered for $\Delta$$V_{BG}$$<$0 (the hole
side) while it vanishes almost completely for $\Delta$$V_{BG}$$>$0
(the electron side). On the contrary, the open-triangle data,
which were obtained with the EC grounded at B, show a significant
reduction of the conductance in the hole side while the
conductance in the electron side recovers the value without
grounded ECs. The reduction of the conductance occurs as the
conducting carriers are leaked to ground before they reach the
drain (D) where the current detector A is located [Fig.
\ref{fig:sample}(a)]. Thus, the result in Fig. \ref{fig:datap}(a)
clearly indicates that the hole-like and the electron-like
carriers in the edge states of QH plateaus circulate in the
opposite directions with each other. To further confirm our
reasoning, we measured $G$ in a negative magnetic field (-9 T) as
shown in Fig. \ref{fig:datan}(a), with the contact resistance
being subtracted. With no grounded ECs, the open-circle data show
almost the same feature as the corresponding open-circle data in
Fig. \ref{fig:datap}(a), along with distorted QH plateaus in the
electron side. However, the open square (triangle) data nearly
coincide with the open triangle (square) data in Fig.
\ref{fig:datap}(a). This indicates that the reduction and recovery
of $G$ depend on the position of the grounded EC, the carrier
type, and the magnetic field direction.

The observed reduction and recovery of the conductance can be
explained by assuming that the chiral property of the QH edge
states in graphene is the same as the one in an ordinary 2D
electron gas.~\cite{Halperin82,Buttiker88} As illustrated in Fig.
\ref{fig:sample}(c), holes (electrons) in graphene in QH plateaus
flow along the upper side of the sample between the source (S) and
the drain (D) in a positive (negative) magnetic field, which
corresponds to the \textit{clockwise} QH edge state. For
simplicity, schematics of the QH edge state circulation are
illustrated in the figures only for the hole side at a positive
$V_{SD}$. Consequently, with no grounded ECs, charge carriers are
not impeded until they reach the drain D, so that $G$ becomes the
Hall conductance as shown by the open-circle symbols in Figs.
\ref{fig:datap}(a) and
\ref{fig:datan}(a).~\cite{Buttiker88,Williams07,Ozyilmaz07,Abanin08,Williams08}
On the other hand, if a grounded EC is placed in the way of the QH
edge state circulation between S and D, electric charges are
diverted to ground before they reach D. The consequent vanishing
of current leads to the reduction of the conductance, as
illustrated in Figs. \ref{fig:datap}(c) and \ref{fig:datan}(b) for
the \textit{clockwise} and \textit{counterclockwise} QH edge
states, respectively. By contrast, if an EC is grounded in the
opposite side of the sample in the circulation path, the current
is not drained to ground before reaching the drain D, resulting in
the recovery of the conductance as illustrated in Figs.
\ref{fig:datap}(b) and \ref{fig:datan}(c).

Up to this point, we assume that conducting carriers are always
emitted from the source S, i.e., a positive (negative) $V_{SD}$
for hole (electron) carriers. But, if hole (electron) carriers are
emitted from both D and the grounded EC for a negative (positive)
$V_{SD}$, the conductance reduction occurs as the same amount of
holes (electrons) enter into and leave from D simultaneously with
vanishing current $I$. Thus, the results in Figs.
\ref{fig:datap}(a) and \ref{fig:datan}(a) directly confirm that
the QH edge states in graphene also behave in the same way as
those in an ordinary 2D electron gas, whose chiral direction
depends on the carrier type and the magnetic field
direction.~\cite{Halperin82,Buttiker88}

The imperfect reduction and recovery of $G$ are seen in Figs.
\ref{fig:datap}(a) and \ref{fig:datan}(a) for
$\Delta$$V_{BG}$$>$15 V and $<$-15 V, which suggest that the
coupling strength between the electric contacts and the QH edge
states is not uniform for different filling
factors.~\cite{Krstic08} It is because the QH edge states with
smaller filling factors circulate closer to the edge of a
sample,~\cite{Halperin82,Buttiker88} with stronger coupling with
the electric contacts near the edge. Therefore, a fraction of
carriers in the QH edge states with higher filling factors of 6
and 10 either reach the drain D by passing the grounded EC in the
way of the carrier circulation (corresponding to the imperfect
reduction) or are drained at the grounded EC in the path after
passing the drain D (corresponding to the imperfect recovery). The
coupling is expected to be stronger for a larger contact area of
the ECs, which explains why the conductance 2$G_0$ shows the full
recovery but with the imperfect reduction.

In summary, we made simple two-terminal conductance measurements
to clarify the chiral properties of the QH edge states in
graphene, which is important to design delicate functionalities in
graphene-based quantum devices. Experimental findings directly
support that the QH edge states in graphene are chiral in the same
way as those in an ordinary 2D electron gas in low magnetic
fields~\cite{Halperin82,Buttiker88,Ki09} ($\sim$9 T). In addition,
from the imperfect reduction and recovery observed in the
experiment, we demonstrate that the coupling strength between the
QH edge states and the electric contacts varies depending on the
filling factors as well as the contact area.~\cite{Krstic08} This
simple two-terminal conductance-measurement scheme can be
conveniently adopted to study the chiral direction of the QH edge
state of zero filling factor in graphene in a stronger magnetic
field~\cite{Zhang06,Abanin07,Jiang07,Checkelsky08,Checkelsky09}
($\gtrsim$25 T), to clarify the quantum state between the two
contradicting models leading to the QH ferromagnets and the QH
insulators.~\cite{Abanin07-1,Gusynin08}

This work was supported by Acceleration Research Grant
(R17-2008-007-01001-0) by Korea Science and Engineering
Foundation.


\begin{thebibliography}{}

\bibitem{Klitzing80} K. v. Klitzing, G. Dorda, and M. Pepper,
Phys. Rev. Lett. {\bf 45}, 494 (1980).

\bibitem{Halperin82} B. I. Halperin, Phys. Rev. B {\bf 25}, 2185
(1982).

\bibitem{Buttiker88} M. B\"{u}ttiker, Phys. Rev. B {\bf 38}, 9375 (1988).

\bibitem{Ji03} Y. Ji, Y. Chung, D. Sprinzak, M. Heiblum, D. Mahalu,
and H. Shtrikman, Nature (London) {\bf 422}, 415 (2003).

\bibitem{Neto09} For the recent review, see A. H. Castro Neto, F.
Guinea, N. M. R. Peres, K. S. Novoselov, and A. K. Geim, Rev. Mod.
Phys. {\bf 81}, 109 (2009).

\bibitem{Abanin07-2} D. A. Abanin and L. S. Levitov, Science {\bf
317}, 641 (2007).

\bibitem{Williams07} J. R. Williams, L. DiCarlo, and C. M. Marcus,
Science {\bf 317}, 638 (2007).

\bibitem{Ozyilmaz07} B. \"{O}zyilmaz, P. Jarillo-Herrero, D. Efetov,
D. A. Abanin, L. S. Levitov, and P. Kim, Phys. Rev. Lett. {\bf 99},
166804 (2007).

\bibitem{Zhang06} Y. Zhang, Z. Jiang, J. P. Small, M. S. Purewal, Y. W.
Tan, M. Fazlollahi, J. D. Chudow, J. A. Jaszczak, H. L. Stormer,
and P. Kim, Phys. Rev. Lett. {\bf 96}, 136806 (2006).

\bibitem{Abanin07} D. A. Abanin, K. S. Novoselov, U. Zeitler, P. A.
Lee, A. K. Geim, and L. S. Levitov, Phys. Rev. Lett. {\bf 98},
196806 (2007).

\bibitem{Jiang07} Z. Jiang, Y. Zhang, H. L. Stormer, and P. Kim,
Phys. Rev. Lett. {\bf 99}, 106802 (2007).

\bibitem{Checkelsky08} J. G. Checkelsky, L. Li, and N. P.
Ong, Phys. Rev. Lett. {\bf 100}, 206801 (2008).

\bibitem{Checkelsky09} J. G. Checkelsky, L. Li, and N. P. Ong, Phys.
Rev. B {\bf 79}, 115434 (2009).

\bibitem{Abanin07-1} D. A. Abanin, P. A. Lee, L. S. Levitov, Solid
State Commun. {\bf 143}, 77 (2007).

\bibitem{Gusynin08} For a review, see V. P. Gusynin, V. A. Miransky, S.
G. Sharapov, and I. A. Shovkovy, Low Temp. Phys. {\bf 34}, 778
(2008).

\bibitem{Ki09} In the recent study on the bipolar p-n-p junction of
graphene, the chiral-direction-dependent asymmetric Hall resistance
was observed. D. K. Ki and H. J. Lee, e-print arXiv:0903.0213.

\bibitem{Krstic08} V. Krstic, D. Obergfell, S. Hansel, Geert L. J. A.
Rikken, J. H. Blokland, M. S. Ferreira, and S. Roth, Nano. Lett.
{\bf 8}, 1700-1703 (2008).

\bibitem{Novoselov04} K. S. Novoselov, A. K. Geim, S. V. Morozov, D.
Jiang, Y. Zhang, S. V. Dubonos, I. V. Grigorieva, and A. A. Firsov,
Science {\bf 306}, 666-669 (2004).

\bibitem{Ki08} D. K. Ki, D. Jeong, J. H. Choi, H. J. Lee, and K. S.
Park, Phys. Rev. B {\bf 78}, 125409 (2008).

\bibitem{Farmer09} D. B. Farmer, R. Golizadeh-Mojarad, V.
Perebeinos, Y. M. Lin, G. S. Tulevski, J. C. Tsang, and P.
Avouris, Nano Lett. {\bf 9}, 388 (2009).

\bibitem{Huard08} B. Huard, N. Stander, J. A. Sulpizio, and D.
Goldhaber-Gordon, Phys. Rev. B {\bf 78}, 121402(R) (2008).

\bibitem{Abanin08} D. A. Abanin and L. S. Levitov, Phys. Rev. B {\bf
78}, 035416 (2008).

\bibitem{Williams08} J. R. Williams, D. A. Abanin, L. DiCarlo, L. S.
Levitov, and C. M. Marcus, e-print arXiv:0810.3397.

\end{thebibliography}
\end{document}